\documentclass[ALICE,manyauthors]{cernphprep}

\newcommand{\s}{\ensuremath{\sqrt{s}}\xspace}
\newcommand{\GeVc}{\ensuremath{\textrm{GeV}/c}\xspace}
\newcommand{\pt}{\ensuremath{p_{\rm T}}\xspace}

\newcommand{\ptrange}[2]{$#1<\pt<#2~\textrm{GeV}/c$}
\newcommand{\dEdx}{\ensuremath{\textrm{d}E/\textrm{d}x}\xspace}
\newcommand{\sNN}{\ensuremath{\sqrt{s_{\rm NN}}~}}
\newcommand{\RpPb}{\ensuremath{R_{\rm pPb}}\xspace}
\newcommand{\RpA}{\ensuremath{R_{\rm pA}}~}

\newcommand{\pPb}{p--Pb~}

\newcommand{\sigmaTPC}{\ensuremath{\sigma_{\textrm{TPC}}}\xspace}
\newcommand{\sigmaTOF}{\ensuremath{\sigma_{\textrm{TOF}}}\xspace}
\newcommand{\nsigmaTPC}{\ensuremath{n_{\sigma}^{\textrm{TPC}}}\xspace}
\newcommand{\nsigmaTOF}{\ensuremath{n_{\sigma}^{\textrm{TOF}}}\xspace}
 
\newcommand{\Ep}{\ensuremath{E/p}} 
\newcommand{\sqpA}{\ensuremath{\sqrt{s_{\rm NN}}} = 5.02~TeV\xspace}
\newcommand{\smb}{\ensuremath{\sigma_{\rm MB}^{\rm V0}}\xspace}
\newcommand{\smbeq}{\sigma_{\rm MB}^{\rm V0}}

\usepackage{color}
\usepackage[comma,square,numbers,sort&compress]{natbib}
\usepackage{siunitx}
\usepackage{hyperref}

\begin{document}

%%%%%%%%%%%%%%%  Title page %%%%%%%%%%%%%%%%%%%%%%%%
\begin{titlepage}
\PHyear{2015}
\PHnumber{262}      % required, will be obtained from PH
\PHdate{17 September}  % required, will be obtained from PH
%

%%% Put your own title + short title here:
\title{ Measurement of electrons from heavy-flavour hadron decays \\
 in \pPb collisions at $\mathbf{\sNN}$ = 5.02 TeV }
\ShortTitle{Measurement of electrons from heavy-flavour hadron decays in \pPb collisions}   % appears on right page headers 

%%% Do not change the next lines
\Collaboration{ALICE Collaboration\thanks{See Appendix~\ref{app:collab} for the list of collaboration members}}
\ShortAuthor{ALICE Collaboration} % appears on left page headers, do not change

\begin{abstract}
The production of electrons from heavy-flavour hadron decays was measured as a function of transverse momentum (\pt) in minimum-bias \pPb collisions at \sqpA using the ALICE detector at the LHC. The measurement covers the \pt interval \ptrange{0.5}{12} and the rapidity range $-1.065 < y_{\rm cms} < 0.135$ in the centre-of-mass reference frame. The contribution of electrons from background sources was subtracted using an invariant mass approach.
The nuclear modification factor \RpPb was calculated by comparing the \pt-differential invariant cross section in \pPb collisions to a pp reference at the same centre-of-mass energy, which was obtained by interpolating measurements at \s~=~2.76~TeV and \s~=~7~TeV. The \RpPb is consistent with unity within uncertainties of about 25\%, which become larger  for \pt below 1~\GeVc. 
The measurement shows that heavy-flavour production is consistent with binary scaling, so that a suppression in the high-\pt yield in Pb--Pb collisions has to be attributed to effects induced by the hot medium produced in the final state. The data in \pPb collisions are described by recent model calculations that include cold nuclear matter effects.

\end{abstract}
\end{titlepage}
\setcounter{page}{2}

\section{Introduction}

The Quark-Gluon Plasma (QGP)~\cite{Bjorken:1982qr,Karsch:2001cy}, a colour-deconfined state of strongly-interacting matter, is predicted to exist at high temperature according to lattice Quantum Chromodynamics (QCD) calculations~\cite{Borsanyi:2013bia}. These conditions can be reached in ultra-relativistic heavy-ion collisions ~\cite{Arsene:2004fa,Back:2004je,Adams:2005dq,Adcox:2004mh,Aamodt:2010pb,Aamodt:2010pa,Aamodt:2010jd}. Charm and beauty (heavy-flavour) quarks are mostly produced in initial hard scattering processes on a very short time scale, shorter than the formation time of the QGP medium~\cite{Liu:2012ax}, and thus experience the full temporal and spatial evolution of the collision.  While interacting with the QGP medium, heavy quarks lose energy via elastic and radiative processes~\cite{Braaten:1991jj,Baier:1996kr,Wicks:2007am}. Heavy-flavour hadrons are therefore well-suited probes to study the properties of the QGP. The effect of energy loss on heavy-flavour production can be characterised via the nuclear modification factor ($R_{\rm AA}$) of heavy-flavour hadrons. The $R_{\rm AA}$ is defined as the ratio of the heavy-flavour hadron yield in nucleus--nucleus (A--A) collisions to that in proton--proton (pp) collisions scaled by the average number of binary nucleon--nucleon collisions. The $R_{\rm AA}$ is studied differentially as a function of transverse momentum (\pt), rapidity ($y$) and collision centrality.
It was measured at the Relativistic Heavy Ion Collider (RHIC)~\cite{Adare:2006nq,Abelev:2006db,Adare:2012px,Adamczyk:2014uip} and at the Large Hadron Collider (LHC)~\cite{Abelev:2012qh,Chatrchyan:2012np,Adam:2015rba,Adam:2015nna}. At RHIC, in central Au--Au collisions at \sNN = 200~GeV the $R_{\rm AA}$ of charmed mesons and of electrons from heavy-flavour hadron decays shows that their production is strongly suppressed by a factor of about 5 for \pt $>$ 3 \GeVc at mid-rapidity. For the most central Pb--Pb collisions at \sNN = 2.76~TeV at the LHC, a suppression by a factor of 5--6 is observed for charmed mesons for \pt $>$ 5 \GeVc at mid-rapidity~\cite{Adam:2015nna}. 

The interpretation of the measurements in A--A collisions requires the study of heavy-flavour production in p--A collisions, which provides access to cold nuclear matter (CNM) effects. These effects are not related to the formation of a colour-deconfined medium, but are present in case of colliding nuclei (or proton--nucleus). An important CNM effect in the initial state is parton-density shadowing or saturation, which can be described using modified parton distribution functions (PDF) in the nucleus~\cite{Armesto:2006ph} or using the Color Glass Condensate (CGC) effective theory~\cite{Fujii:2006ab}. Further CNM effects include energy loss~\cite{Kang:2014hha} in the initial and final states and a Cronin-like enhancement~\cite{Cronin:1974zm} as a consequence of multiple scatterings~\cite{Sharma:2009hn,Kang:2014hha}.

The influence of the CNM effects can be studied by measuring the nuclear modification factor $R_{\rm pA}$.
Like the $R_{\rm AA}$, the $R_{\rm pA}$ is defined such that it is unity if there are no nuclear effects. For minimum-bias p--A collisions, it can be expressed as~\cite{d'Enterria:2003qs}
\begin{equation}
% \RpA = \frac{1}{<T_{\rm pPb}>}\frac{dN_{\rm pPb}/d\pt}{d\sigma_{\rm pp}/d\pt}
\RpA = \frac{1}{A}\frac{{\rm d}\sigma_{\rm pA}/{\rm d}\pt}{{\rm d}\sigma_{\rm pp}/{\rm d}\pt}~,
\end{equation}
where ${\rm d}\sigma_{\rm pA}/{\rm d}\pt$ and ${\rm d}\sigma_{\rm pp}/{\rm d}\pt$ are the \pt-differential production cross sections of a given particle species in p--A and pp collisions, respectively, and $A$ is the number of nucleons in the nucleus.

Cold nuclear matter effects were recently investigated at the RHIC and the LHC~\cite{Adams:2004fc,Adare:2012bv,Adare:2012qf,Adare:2013ezl,Adare:2013xlp,Adamczyk:2013poh,Adare:2014iwg,Abelev:2014zpa,Abelev:2014oea,Adam:2015iga,Aad:2015ddl,Chatrchyan:2013nza,Khachatryan:2015uja,Adare:2012yxa,Adare:2013lkk,Abelev:2014hha}. At RHIC, the nuclear modification factor of electrons from heavy-flavour hadron decays in central d--Au collisions (0--20\%) at \sNN = 200~GeV is larger than unity at mid-rapidity in the transverse momentum interval \ptrange{1.5}{5}~\cite{Adare:2012yxa}. The corresponding measurement for muons from heavy-flavour hadron decays in central d--Au collisions shows a suppression at forward rapidity and an enhancement at backward rapidity~\cite{Adare:2013lkk}. Theoretical models that include the modification of the PDF in the nucleus can neither explain the enhancement nor the large difference between forward and backward rapidity. Possible explanations include the Cronin-like enhancement~\cite{Cronin:1974zm} due to radial flow of heavy mesons \cite{Sickles:2013yna}.  At the LHC, the \pt-differential nuclear modification factor \RpPb of D mesons measured in \pPb collisions at \sqpA ~\cite{Abelev:2014hha} is consistent with unity for \pt\ $>$ 1~\GeVc\ and is described by theoretical calculations that include gluon saturation effects. Both at RHIC and at the LHC, the p/d--A measurements indicate that initial-state effects alone cannot explain the strong suppression seen at high-\pt in nucleus--nucleus collisions. 

In this Letter, the \pt-differential invariant cross section and the nuclear modification factor \RpPb of electrons from heavy-flavour hadron decays measured in minimum-bias \pPb collisions at \sqpA with ALICE at the LHC are presented. The measurement covers the rapidity range $-1.065 < y_{\rm cms} < 0.135$ in the centre-of-mass system (cms) for electrons with transverse momentum \ptrange{0.5}{12}. 
This rapidity coverage results from the same rigidity of the p and Pb beams at the LHC, leading to a rapidity shift of $|y_{\rm NN}|$ = 0.465 between the nucleon--nucleon cms and the laboratory reference frame, in the direction of the p beam.
At low \pt, the measurement probes the production of charm-hadron decays~\cite{Abelev:2012sca}, providing sensitivity to the gluon PDF in the regime of Bjorken-$x$ of the order of $10^{-4}$~\cite{Dainese:2003zu}, where a substantial shadowing effect is expected~\cite{Eskola:2009uj}.

To obtain the nuclear modification factor \RpPb of electrons from heavy-flavour hadron decays, the \pt-differential invariant cross section in \pPb collisions at \sqpA was compared to a pp reference multiplied by 208, the Pb mass number. The pp reference was obtained by interpolating the $p_{\rm T}$-differential cross section measurements at $\s=2.76$~TeV and 7~TeV.

The Letter is organised as follows. The experimental apparatus, data sample and event selection are described in Section~\ref{Exp}. The electron reconstruction strategy and the pp reference spectrum are explained in Sections~\ref{Ana} and~\ref{ppref}, respectively. The measured \pt-differential invariant cross section, the nuclear modification factor \RpPb of electrons from heavy-flavour hadron decays and comparison of \RpPb to model calculations are reported in Section~\ref{Res}.

\section{Experimental apparatus, data sample and event selection}\label{Exp}

A detailed description of the ALICE apparatus can be found in~\cite{Aamodt:2008zz,Abelev:2014ffa}. Electrons are reconstructed at mid-rapidity using the central barrel detectors (described below)
located inside a solenoid magnet, which generates a magnetic field $\mathrm{B}=0.5$~T along the beam direction.

The Inner Tracking System (ITS), the closest detector to the interaction point, includes six cylindrical layers of silicon detectors with three different technologies (pixel, drift and strip) at radii between 3.9~cm and 43~cm with a pseudorapidity coverage in the laboratory reference frame in the full azimuth between $|\eta_{\rm lab}|<$~2.0 at small radii and $|\eta_{\rm lab}|<$~0.9 at large radii~\cite{Aamodt:2008zz,Aamodt:2010aa}. The two innermost layers form the Silicon Pixel Detector (SPD), which plays a key role in primary and secondary vertex reconstruction. At an incident angle perpendicular to the detector surfaces, the total material budget of the ITS corresponds on average to 7.7\% of a radiation length~\cite{Aamodt:2010aa}. The main tracking device in the central barrel is the Time Projection Chamber (TPC)~\cite{Alme:2010ke}, which surrounds the ITS and covers a pseudorapidity range of $|\eta_{\rm lab}|<$ 0.9 in the full azimuth. The track reconstruction proceeds inward from the outer radius of the TPC to the innermost layer of the ITS~\cite{Abelev:2014ffa}. The TPC provides particle identification via the measurement of the specific energy loss d$E$/d$x$. The Time-Of-Flight array (TOF), based on Multi-gap Resistive Plate Chambers, covers the full azimuth and $|\eta_{\rm lab}|<$~0.9 at a radial distance of 3.7~m from the interaction point~\cite{Cortese:2002kf}. Using the particle time-of-flight measurement, electrons can be distinguished from hadrons for \mbox{\pt $\le$ 2.5 \GeVc}. The collision time, used for the calculation of the time-of-flight to the TOF detector, is measured by an array of Cherenkov counters, the T0 detector, located at $+350$~cm and $-70$~cm from the interaction point along the beam direction~\cite{Cortese:2004aa}. The Electromagnetic Calorimeter (EMCal), situated behind the TOF, is a sampling calorimeter based on Shashlik technology~\cite{Cortese:2008zza}. Its geometrical acceptance is 107$^{\circ}$ in azimuth and $|\eta_{\rm lab}|< $ 0.7. In this analysis, the azimuthal angle and $\eta$ coverage were limited to 100$^{\circ}$ and 0.6, respectively, to ensure uniform detector performance. 

The minimum-bias (MB) \pPb data sample used in this analysis was collected in 2013. The trigger condition required a coincidence of signals between the two V0 scintillator hodoscopes, placed on either side of the interaction point at 2.8 $ < \eta_{\rm lab} <$ 5.1 and $-$3.7 $ < \eta_{\rm lab} <$ $-$1.7, synchronised with the passage of bunches from both beams~\cite{Cortese:2004aa}. The background due to interactions of one of the two beams and residual particles in the beam vacuum tube was rejected in the offline event selection by correlating the time information of the V0 detectors with that from the two Zero Degree Calorimeters (ZDC)~\cite{Abelev:2014ffa}, that are located 112.5~m away from the interaction point along the beam pipe, symmetrically on either side. 
The primary vertex was reconstructed with tracks in the ITS and the TPC~\cite{Abelev:2014ffa}. 
Events with a primary vertex located farther than $\pm$~10~cm from the centre of the interaction region along the beam direction were rejected. About 10\% of the events do not fulfil this selection criterion. A sample of 100 million events passed the offline event selection, corresponding to an integrated luminosity \mbox{$L_{\mathrm{int}} = 47.8 \pm 1.6~\mu\mathrm{b}^{-1}$}, given the cross section $\smb = 2.09\pm0.07$~b for the minimum-bias V0 trigger condition~\cite{Abelev:2014epa}. The efficiency for the trigger condition and offline event selection is larger than $99\%$ for non-single-diffractive (NSD) \pPb collisions~\cite{ALICE:2012xs}.

\section{Analysis}\label{Ana}

A combination of electron identification (eID) strategies with different detectors offers the largest \pt reach for the measurement of electrons from heavy-flavour hadron decays. In particular, it ensures that the systematic uncertainties and the hadron contamination are small over the whole transverse momentum range. Throughout the paper, the term `electron' is used for electrons and positrons. The capability of the TPC to identify electrons via specific energy loss \dEdx in the detector was used over the whole momentum range \ptrange{0.5}{12}. However, it is subject to ambiguous identification of hadrons (pions, kaons, protons and deuterons) below 2.5~\GeVc and above 6~\GeVc in transverse momentum. At low transverse momentum (\ptrange{0.5}{2.5}), these ambiguities were resolved by measuring the time-of-flight of the particle from the interaction region to the TOF detector and combining it with the momentum measurement, to determine the particle mass. In the high momentum region (\ptrange{6}{12}), the EMCal was used to reduce the hadron contamination. Electrons are separated from hadrons by calculating the ratio of the energy deposited ($E$) in the EMCal to the momentum ($p$). Since electrons deposit all of their energy in the EMCal, the ratio \Ep\ is around unity for electrons, while the ratio for charged hadrons is much smaller on average.

The selection criteria for charged-particle tracks are similar to those applied in previous analyses measuring the production of electrons from heavy-flavour hadron decays in pp collisions~\cite{Abelev:2014gla,Abelev:2012xe}. In order to have optimal eID performance with the TPC, the analysis was restricted to the pseudorapidity range $|\eta_{\rm lab}|< $ 0.6 in the laboratory frame for electrons with transverse momentum \ptrange{0.5}{12}. 
Up to a \pt of 6 \GeVc, a signal in the innermost layer of the SPD was required in order to reduce the background from photon conversions. In addition, this selection was further constrained by requiring hits in both SPD layers, to reduce the number of incorrect matches between candidate tracks and hits reconstructed in the first layer of the SPD. At high \pt, where the EMCal was used, tracks with hits in either of the SPD layers were selected in order to minimise the effect of dead areas of the first SPD layer within the acceptance region of the EMCal, as in previous analyses~\cite{Abelev:2014gla,Abelev:2012xe}. 

\begin{figure}[\t] 
\centering
  \subfigure[]{% 
    \includegraphics[scale=0.35]{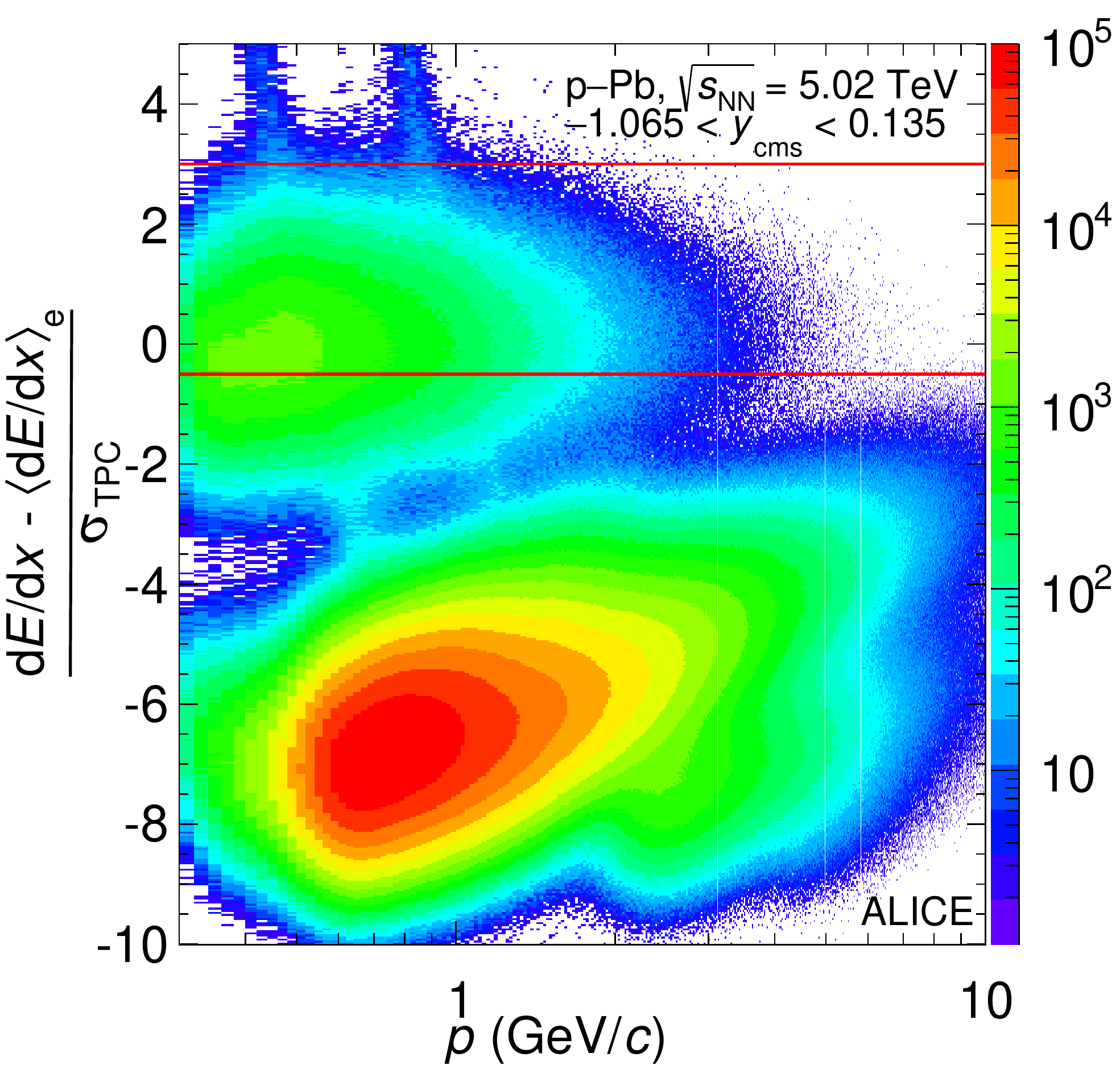} \label{fig::PIDa} } 
  \quad 
  \subfigure[]{% 
    \includegraphics[scale=0.35]{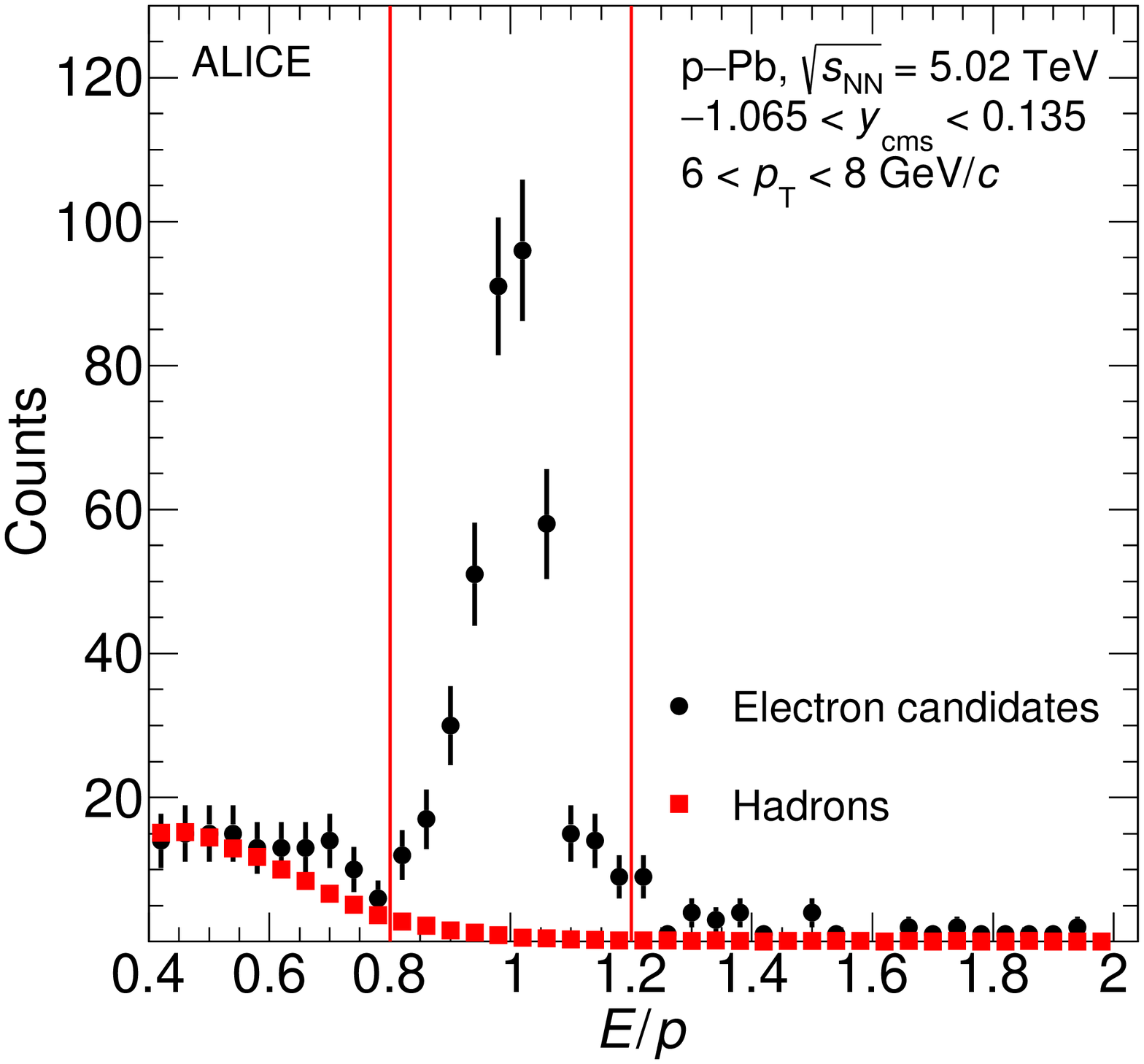} \label{fig::PIDb} } 
\caption{(a): Measured \dEdx in the TPC as function of momentum $p$ expressed as a deviation from the expected energy loss of electrons, normalised by the energy-loss resolution ($\sigmaTPC$) after eID with TOF. The solid lines indicate the $\nsigmaTPC$ selection criteria for the TPC and TOF eID strategy.  (b): \Ep\ distribution of electrons ($-1 < \nsigmaTPC < 3$) and hadrons  ($\nsigmaTPC < -3.5$) in the transverse momentum interval \ptrange{6}{8}. The \Ep\ distribution of hadrons was normalised to that of electrons in the lower \Ep\ range (0.4--0.6), where hadrons dominate. The solid lines indicate the applied electron selection criteria.}
\end{figure}

The electron identification with TPC and TOF was based on the number of standard deviations ($\nsigmaTPC$ or $\nsigmaTOF$) for the specific energy loss and time-of-flight measurements, respectively. The $\ensuremath{n_{\sigma}}$ variable is computed as a difference between the measured signal and the expected one for electrons divided by the energy loss ($\sigmaTPC$) or time-of-flight ($\sigmaTOF$) resolution. The expected signal and resolution originate from parametrisations of the detector signal, which are described in detail in~\cite{Abelev:2014ffa}. In the transverse momentum interval \ptrange{0.5}{2.5}, particles were identified as electrons if they satisfied $-0.5 < \nsigmaTPC < 3$, which yields an identification efficiency of 69\%. In the transverse momentum interval \ptrange{2.5}{6}, a tighter selection criterion of $0 < \nsigmaTPC < 3$ was applied (with an eID efficiency of 50\%) to reduce the hadron contamination at higher transverse momentum. To resolve the aforementioned ambiguities at low transverse momentum (\pt $\le$ \mbox{2.5 \GeVc)}, only tracks with $|\nsigmaTOF| < 3$ were accepted. Figure~\ref{fig::PIDa} shows the measured \dEdx in the TPC with respect to the expected \dEdx for electrons normalised to the expected resolution $\sigmaTPC$ after the eID with TOF. The solid lines indicate the selection criteria used for the transverse momentum interval \ptrange{0.5}{2.5}, indicating that the hadron contamination within the resulting electron candidate sample is small. In the high momentum region (\ptrange{6}{12}), electrons were selected if they satisfied $-1 < \nsigmaTPC < 3$ and $0.8 < E/p < 1.2$ (see Fig.~\ref{fig::PIDb}).

The hadron contamination in the electron candidate sample was determined by parametrising the TPC signal in momentum slices for \mbox{\pt $\le$ 6 \GeVc} as done in previous analyses~\cite{Abelev:2014gla,Abelev:2012xe}. In the transverse momentum interval \ptrange{6}{12}, the \Ep\ distribution for hadrons identified via the specific energy loss measured in the TPC ($\nsigmaTPC < -3.5$)  was normalised in the lower \Ep\ range (0.4--0.6) to the corresponding \Ep\ distribution for identified electrons ($-1 < \nsigmaTPC < 3$) (see Fig.~\ref{fig::PIDb}). The number of hadrons with an \Ep\ ratio between 0.8 and 1.2 was thus determined in momentum slices. The hadron contamination ranged from 2\% at \ptrange{5}{6} to 15\% at \ptrange{10}{12} and was correspondingly subtracted. For \pt~$<$~5~\GeVc, the contamination was found to be negligible.

The resulting electron candidate sample, also referred to as the `inclusive electron sample' in the following, still contains electrons from sources other than heavy-flavour hadron decays. The majority of the remaining background originates from photon conversions in the detector material ($\gamma \rightarrow \rm{e^+ e^-}$) and Dalitz decays of neutral mesons, e.g. $\pi^0 \rightarrow \gamma\ \rm{e^+} \rm{e^-}$ and $\eta \rightarrow \gamma\ \rm{e^+   e^-}$. These electrons are hereafter denoted as `photonic electrons'.  

In previous analyses of electrons from heavy-flavour hadron decays in pp collisions by the ALICE Collaboration, the contribution of electrons  from background sources was estimated via a data-tuned Monte Carlo cocktail and subtracted from the inclusive electron sample~\cite{Abelev:2014gla,Abelev:2012xe}. The pion input to the cocktail was based on pion measurements with ALICE~\cite{Abelev:2012cn,Abelev:2014ypa}, while heavier mesons were implemented via $m_{\rm T}$ scaling \cite{Adare:2009qk}, and photons from hard scattering processes (direct $\gamma$, $\gamma^{\ast}$) were obtained from next-to-leading order (NLO) calculations~\cite{Gordon:1994ut}. The resulting systematic uncertainty of the sum of all background sources was large, in particular at low \pt, where the signal-to-background ratio is small~\cite{Abelev:2014gla,Abelev:2012xe}. In order to reduce this uncertainty, in this analysis an invariant mass technique~\cite{Abelev:2006db} was used to estimate the number of electrons coming from background sources.

\begin{figure}
\centering
  \includegraphics[width=0.45\textwidth]{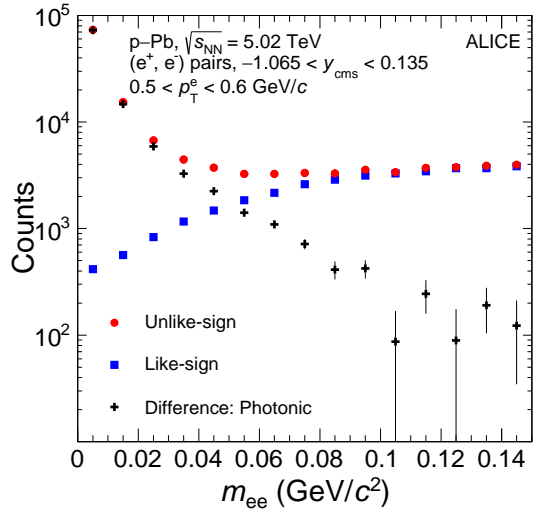}
  \caption{Invariant mass distributions of unlike-sign and like-sign electron pairs for the inclusive electron \pt interval \mbox{\ptrange{0.5}{0.6}}. The difference between the distributions yields the photonic contribution.}
  \label{fig::anglemassPair}
\end{figure}

Photonic electrons are produced in $\rm{e^+ e^-}$ pairs and can thus be identified using an invariant mass technique (photonic method). 
All inclusive electrons were paired with other tracks in the same event passing looser track selection and electron identification criteria (e.g. $-3 < \nsigmaTPC < 3$). Looser selection criteria were applied to increase the efficiency to find the photonic partner.
Figure~\ref{fig::anglemassPair} shows the invariant mass distributions of unlike-sign and like-sign electron pairs for the inclusive electron in the interval \ptrange{0.5}{0.6}. The like-sign distribution estimates the uncorrelated pairs. Subtracting these from the unlike-sign pairs yields the number of electrons with a photonic partner $N_{\rm phot}^{\rm raw}$ (see Fig.~\ref{fig::anglemassPair}). An invariant mass smaller than 0.14 GeV/${c^2}$ was required. According to simulations, the peak around zero in the photonic electron pair distribution is due to photon conversions; the exponential tail to higher values originates from Dalitz decays of neutral mesons. 

The efficiency $\varepsilon_{\rm phot}$ to find photonic electron pairs was estimated using Monte Carlo simulations. A sample of \pPb collisions was generated with HIJING v1.36~\cite{Gyulassy:1994ew}. To increase the statistical precision at high \pt, one $\rm c\overline{c}$ or $\rm b\overline{b}$ pair decaying semileptonically using the generator PYTHIA v6.4.21~\cite{Sjostrand:2006za} with the \mbox{Perugia-0} tune~\cite{Skands:2010ak} was added in each event. The generated particles were propagated through the apparatus using GEANT3~\cite{Brun:1994aa} and a realistic detector response was applied to reproduce the performance of the detector system during data taking period. The simulated transverse momentum distributions of the $\pi^0$ and $\eta$ mesons were weighted to match the measured shapes, where the $\pi^0$ input was based on the measured charged-pion spectra~\cite{Abelev:2013haa,Adam:2016dau} assuming $N_{\pi^0} = 1/2(N_{\pi^+} + N_{\pi^-})$ and the $\eta$ input was derived via $m_{\rm T}$ scaling. The efficiency $\varepsilon_{\rm phot}$ is defined as the fraction of electrons from photonic origin for which the partner could be found within the defined acceptance of the analysis, i.e. the geometrical acceptance of the ALICE apparatus together with the superimposed track selection and electron identification criteria. The efficiency $\varepsilon_{\rm phot}$ increases sharply with \pt from 35\% to 80\% between 0.5 and \mbox{3 \GeVc} and remains at 80\% up to 12 \GeVc. The raw photonic electron distribution $N_{\rm phot}^{\rm raw}$ was then corrected by the efficiency $\varepsilon_{\rm phot}$ as $ N_{\rm phot}(p_{\rm T}) = N_{\rm phot}^{\rm raw}(p_{\rm T})/\varepsilon_{\rm phot}(p_{\rm T})$ and subtracted from the inclusive electron yield to obtain the yield of electrons from heavy-flavour hadron decays. The signal-to-background ratio (ratio of non-photonic to photonic yield) ranges from 0.2 at 0.5 GeV/$c$ to 4 at 10 GeV/$c$.

The remaining electrons are then those from semileptonic heavy-flavour hadron decays ($N_{\rm hfe}^{\rm raw}$), besides a small residual background contribution originating from semileptonic kaon decays and dielectron decays of J/$\psi$ mesons. The latter is the only non-negligible contribution from quarkonia. These contributions were subtracted from the corrected invariant cross section, as described later on in this section.

The $\pt$-differential invariant cross section $\sigma_{\rm hfe}$ of electrons from heavy-flavour hadron decays,  $1/2(\rm{e^+ + e^-})$, was calculated as
\begin{equation}
\frac{1}{2\pi\pt}\frac{{\rm d}^{2}\sigma_{\rm hfe}}{{\rm d}\pt{\rm d}y}=
% E\frac{{\rm d}^{3} \sigma_{\rm hfe}}{{\rm d}p^{3}}=
\frac{1}{2}
\frac{1}{\Delta \varphi \pt^{\rm{centre}}} 
\frac{1}{\Delta y \Delta \pt}
\frac{{\rm c_{\rm unfold}} N_{\rm hfe}^{\rm raw}}{(\epsilon^{\rm geo} \times \epsilon^{\rm reco} \times \epsilon^{\rm eID})}
\frac{\smbeq}{N_{\rm MB}},
\label{eq:cross_section}
\end{equation}
where $\pt^{\rm{centre}}$ are the centres of the \pt\ bins with widths $\Delta\pt$, and $\Delta \varphi$ and $\Delta y$ denote the geometrical acceptance in azimuth and rapidity to which the analysis was restricted, respectively. $N_{\rm MB}$ is the number of events that pass the selection criteria described in Section~\ref{Exp} and \smb is the \pPb cross section for the minimum-bias V0 trigger condition.
The raw spectrum of electrons from heavy-flavour hadron decays ($N_{\rm hfe}^{\rm  raw}$) was corrected for the acceptance of the detectors in the selected geometrical region of the analysis ($\epsilon^{\rm geo}$), the track reconstruction and selection efficiency ($\epsilon^{\rm reco}$), and the eID efficiency ($\epsilon^{\rm eID}$). These corrections were computed using the aforementioned Monte Carlo simulations. Only the efficiency of the TPC electron identification selection criterion for $\pt < 6~\GeVc$ was determined using a data-driven approach based on the \nsigmaTPC distribution~\cite{Abelev:2012xe}. The measurement of the electron \pt is affected by the finite momentum resolution and by electron energy loss due to bremsstrahlung in the detector material~\cite{Aamodt:2008zz}, which is not corrected for in the track reconstruction algorithm. These effects distort the shape of the \pt distribution, which falls steeply with increasing momentum. To determine this correction ($c_{\rm unfold}$), an iterative unfolding procedure based on Bayes' theorem was applied \cite{D'Agostini:1999cm,Grosse-Oetringhaus:2009mka}. 

The aforementioned residual background contributions, electrons from semileptonic kaon decays and dielectron decays of J/$\psi$ mesons, were estimated as an invariant cross section with Monte Carlo simulations and found to be less than 3\% per \pt bin and subtracted from the corrected invariant cross section of non-photonic electrons. More specifically, the contribution from J/$\psi$ mesons was implemented by using a parametrisation for pp collisions based on the interpolation of J/$\psi$ measurements from RHIC at \s~=~200~GeV, Tevatron at \s~=~1.96~TeV, and the LHC at \s~=~7~TeV according to~\cite{Bossu:2011qe}. Decays of J/$\psi$ mesons within $|y_{\rm lab}|< $ 1.0 were considered. The parametrisation and its associated systematic uncertainty were scaled from pp to \pPb collisions assuming binary collision scaling. Potential deviations from binary collision scaling were considered by assigning a 50\% systematic uncertainty on the normalisation.
The parametrisation with its uncertainties used as input for the Monte Carlo simulations is consistent with the measured J/$\psi$ cross section in \pPb collisions~\cite{Adam:2015iga}.

The systematic uncertainties were estimated as a function of \pt by repeating the analysis with different selection criteria. The systematic uncertainties were evaluated for the spectrum obtained after the subtraction of the photonic yield $N_{\rm phot}$ from the inclusive spectrum and before removing the remaining background contributions originating from semileptonic kaon decays and dielectron decays of J/$\psi$ mesons. The sources of systematic uncertainty for the inclusive analysis and the determination of the electron background are listed in Table~\ref{tab::syst}.

The systematic uncertainties for tracking and eID are \pt dependent due to the usage of the various detectors in the different momentum intervals. The latter also includes the uncertainties due to the determination of the hadron contamination. The 3\% systematic uncertainty for the matching between ITS and TPC was taken from~\cite{Abelev:2014dsa}, where the matching efficiency of charged particles in data was compared to Monte Carlo simulations. The uncertainty of the TOF-TPC matching efficiency was estimated by comparing the matching efficiency in data and Monte Carlo simulations using electrons from photon conversions, which were identified via topological cuts. The uncertainty amounts to 3\%. The TPC-EMCal matching uncertainty was assigned to be 1\%, as determined by varying the size of the matching window in $\eta$ and azimuth $\varphi$ for charged-particle tracks that were extrapolated to the calorimeter. The resulting matching uncertainties were combined in quadrature for the various \pt intervals shown in Table~\ref{tab::syst}. 

The listed uncertainties for the photonic method include the uncertainties on eID and tracking. In addition, the Monte Carlo sample was divided into two halves. The first was treated as real data and the second was used to correct the resulting spectrum. Deviations from the expected \pt spectrum of electrons from heavy-flavour hadron decays resulted in a 2\% systematic uncertainty for \mbox{\pt $\le$ 6 GeV/$c$ and 4\%} above. The uncertainty on the re-weighting of the $\pi^0$- and $\eta$-meson \pt distributions in Monte Carlo simulations was estimated by changing the weights by $\pm$ 10\%. The variation yielded a 2\% uncertainty for \mbox{\pt $\le$ 2.5 GeV/$c$} on the \pt-differential invariant cross section of electrons from heavy-flavour hadron decays. This source of uncertainty is negligible at higher \pt. The invariant mass technique gives a systematic uncertainty smaller by a factor of $\ge$ 4 and of about 1.4 for \mbox{\pt $\le$ 1 \GeVc} and \ptrange{3}{12}, respectively, compared to the one of the cocktail subtraction method~\cite{Abelev:2012xe}. The reduction in uncertainty, in particular at low \pt, proves the advantage of using the invariant mass technique for the estimation of electrons from background sources.

The uncertainty of the \pt unfolding procedure was determined by employing an alternative unfolding method (matrix inversion) and, as described in~\cite{Abelev:2012xe}, by correcting the data with two different Monte Carlo samples corresponding to different \pt distributions. In addition to the aforementioned signal-enhanced Monte Carlo sample, a minimum-bias sample was used. The comparison of the resulting \pt spectra revealed an uncertainty of 1\% for \pt $\leq$ \mbox{6~\GeVc,} and smaller than 1\% above 6~\GeVc. The systematic uncertainties of the heavy-flavour electron yield due to the subtraction of the remaining background originating from semileptonic kaon decays and dielectron decays from  J/$\psi$ mesons are smaller than 0.5\%. This was estimated by changing the particle yields by $\pm$ 50\% and $\pm$ 100\% for the J/$\psi$ meson and the semileptonic kaon decays, respectively. 

The individual sources of systematic uncertainties are uncorrelated. Therefore, they were added in quadrature to give a total systematic uncertainty ranging from 5.8\% to 16.4\% depending on the \pt bin. The normalisation uncertainty on the luminosity is of 3.7\% ~\cite{Abelev:2014epa}.

\begin{table}[h]
\centering
\scalebox{0.9}{
\begin{tabular}{|l|c|c|c|}

\hline
Variable      &\ptrange{0.5}{2.5}& \ptrange{2.5}{6} & \ptrange{6}{12}\\
\hline
Tracking 			& 4.3\%           &  2.2\%        & 3\%   \\
Matching                 & 4.2\%           & 3\%            & 3.2\%     \\
 \hline

eID                           & 3.6\%           & 3.6\%  & \begin{tabular}{@{}l@{}}~~3.2\% (6--8\,\GeVc) \\ ~~5.1\% (8--10\,\GeVc) \\15.1\% (10--12\,\GeVc)\end{tabular} \\        
\hline
\begin{tabular}{@{}l@{}} Photonic \\ method  \end{tabular}              & \begin{tabular}{@{}l@{}}6.9\% (0.5--1\,\GeVc) \\ 3.7\% (1--2.5\,\GeVc) \end{tabular}          &  2.4\%        & 4.5\%   \\ \hline
Unfolding                  &  1\%            &  1\%          &  $<$1\% \\ \hline \hline     
Total            &     \begin{tabular}{@{}l@{}}9.9\% (0.5--1\,\GeVc) \\ 8.0\% (1--2.5\,\GeVc) \end{tabular}           &  5.8\%  	  &  \begin{tabular}{@{}l@{}}~~7.1\% (6--8\,\GeVc) \\ ~~8.1\% (8--10\,\GeVc) \\16.4\% (10--12\,\GeVc)\end{tabular} \\                              
\hline
\end{tabular}
}
\caption{Systematic uncertainties for the different momentum intervals.}
\label{tab::syst}
\end{table}

\begin{figure}[\hbt]
\centering
  \includegraphics[scale=0.45]{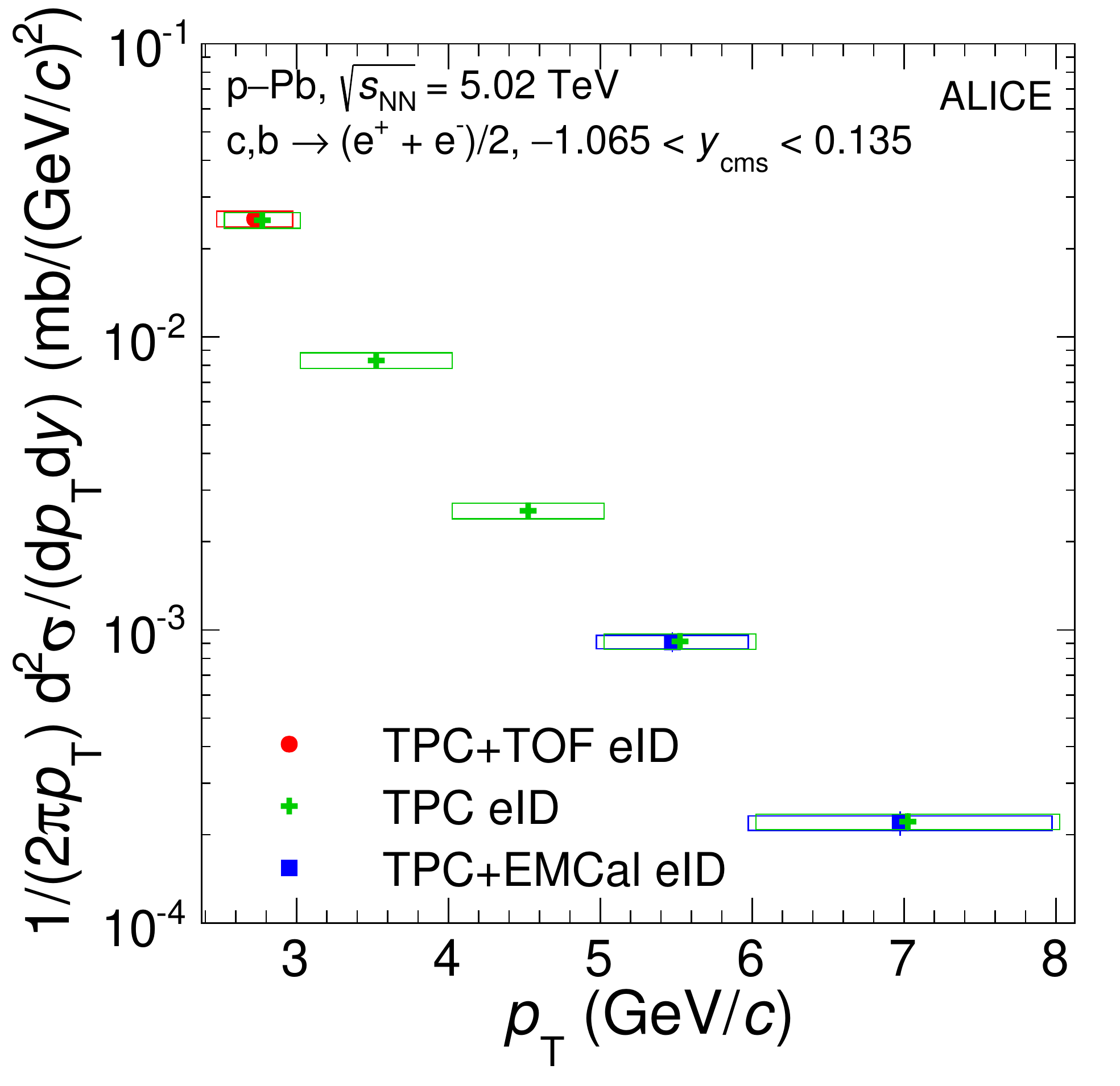}
  \caption{The \pt-differential invariant cross section of electrons from heavy-flavour hadron decays in minimum-bias \pPb collisions at \mbox{\sqpA}, comparing the results of the eID strategies in the two transition regions at 2.5 and 6 \GeVc. The centre values are slightly shifted along the \pt-axis in the transition regions for better visibility. The results agree within 1\%. Details on the eID strategies can be found in the text.}
\label{fig::comparison}
 \end{figure}

Figure~\ref{fig::comparison} shows the interval \ptrange{2.5}{8} of the \pt-differential invariant cross section of electrons from heavy-flavour hadron decays in minimum-bias \pPb collisions at \mbox{\sqpA}, comparing the results of the various eID strategies in the two transition regions at 2.5 \GeVc and 6 \GeVc. A consistency within 1\% is found.

\section{pp reference}\label{ppref}

In order to calculate the nuclear modification factor $R_{\rm pPb}$, a reference cross section for pp collisions at the same centre-of-mass energy is needed. Since pp data at $\s = 5.02$~TeV are currently not available, the reference was obtained by interpolating the \pt-differential cross sections of electrons from heavy-flavour hadron decays measured in pp collisions at $\s=2.76$~TeV and at $\s=7$~TeV~\cite{Abelev:2014gla,Abelev:2012xe}. The analysis described in this paper requires a reference in the interval \ptrange{0.5}{12}. While the $\s=2.76$~TeV analysis was carried out in this \pt range, the $\s=7$~TeV measurement is limited to the \pt interval \ptrange{0.5}{8}. Thus, to extend the \pt interval up to 12~\GeVc a measurement by the ATLAS Collaboration in the \pt interval \ptrange{7}{12} was used~\cite{Aad:2011rr}. The published ATLAS measurement, d$\sigma$/d\pt, was divided by 1/(2$\pi\pt^{\rm{centre}}\Delta y$), where $\pt^{\rm{centre}}$ denotes the central values of the \pt\ bins, and $\Delta y$ the rapidity range covered by the measurement. In the overlap  interval \ptrange{7}{8} the mbox{ALICE} and ATLAS measurements, which agree within uncertainties, were combined as a weighted average. The inverse quadratic sum of statistical and systematic uncertainties of the two spectra were used as weights. Perturbative QCD (pQCD) calculations at fixed order with next-to-leading-log (FONLL) resummation~\cite{Cacciari:1998it,Cacciari:2001td,Cacciari:2012ny} describe all aforementioned pp results~\cite{Abelev:2014gla,Abelev:2012xe} within experimental and theoretical uncertainties. The pp references are measured in a symmetric rapidity window ($|y_{\textrm{cms}}|<0.8$ at $\s=2.76$~TeV and $|y_{\textrm{cms}}|<0.5$ at $\s=7$~TeV). The effect due to the different asymmetric rapidity window in this analysis was estimated with FONLL and is much smaller than the systematic uncertainties of the data, therefore is was neglected.

An assumption about the $\sqrt{s}$ dependence of the heavy-flavour production cross sections is required for the interpolation. Calculations based on pQCD are consistent with a power-law scaling of the heavy-flavour production cross section with $\sqrt{s}$~\cite{Abelev:2012vra}. Therefore, this scaling was used to calculate the interpolated data points. The statistical uncertainties of the spectra at $\s=2.76$~TeV and $\s=7$~TeV were added in quadrature with weights according to the $\sqrt{s}$ interpolation. The weighted correlated systematic uncertainties (tracking, matching and eID) of the spectra at $\s=2.76$~TeV and $\s=7$~TeV were added linearly, while the weighted uncorrelated uncertainties (ITS layer conditions, unfolding and cocktail systematics) were added in quadrature. The weights were determined according to the $\sqrt{s}$ interpolation. The uncorrelated and correlated uncertainties were then added in quadrature. 

The systematic uncertainty of the bin-by-bin interpolation procedure was added in quadrature to the previous ones. It was estimated by using a linear or exponential dependence on $\sqrt{s}$ instead of a power law. The ratios of the resulting \pt spectra to the baseline pp reference were used to estimate a systematic uncertainty of $^{+\phantom{a}5}_{-10}$\%.

The resulting pp reference cross section is well described by FONLL calculations. 
The systematic uncertainties of the normalisations related to the determination of the minimum-bias nucleon--nucleon cross sections of the input spectra were likewise interpolated, yielding a normalisation uncertainty of 2.3\% for the pp reference spectrum, assuming that they are uncorrelated.

\section{Results}\label{Res}

\begin{figure}[\t]
\centering
  \includegraphics[scale=0.45]{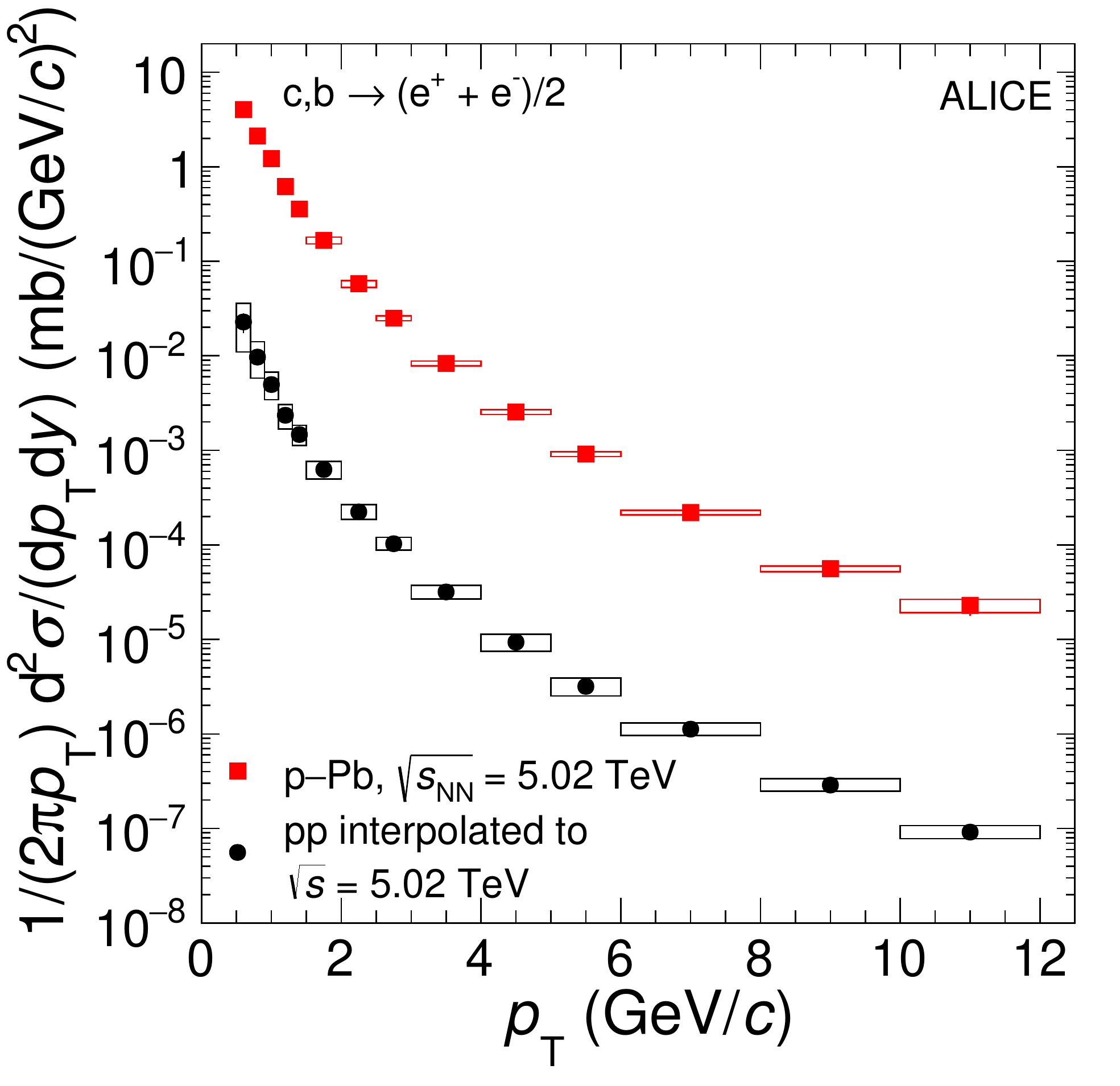}
  \caption{The \pt-differential invariant cross section of electrons from heavy-flavour hadron decays in minimum-bias \pPb collisions at \mbox{\sqpA}. The pp reference obtained via the interpolation method is shown, not scaled by $A$, for comparison. The statistical uncertainties are indicated for both spectra by error bars, the systematic uncertainties are shown as boxes.}
\label{fig::hfept}
 \end{figure}

The \pt-differential invariant cross section of electrons from heavy-flavour hadron decays in the rapidity range $-1.065 < y_{\rm cms} < 0.135$ for \pPb collisions at \mbox{\sqpA} is shown in Fig.~\ref{fig::hfept} and compared with the pp reference cross section. The vertical bars represent the statistical uncertainties, while the boxes indicate the systematic uncertainties. The systematic uncertainties of the \pPb cross section are smaller than those of the pp cross section, in particular at low transverse momentum, mainly as a consequence of the estimation of the electron background via the invariant mass technique. For the pp analysis, the background was subtracted via the cocktail method. At low \pt, the electrons mainly originate from charm-hadron decays, while for \pt $\ge$ 4 GeV/$c$ beauty-hadron decays are the dominant source in pp collisions~\cite{Abelev:2012sca}.

\begin{figure}[\t]
\centering
  \includegraphics[scale=0.45]{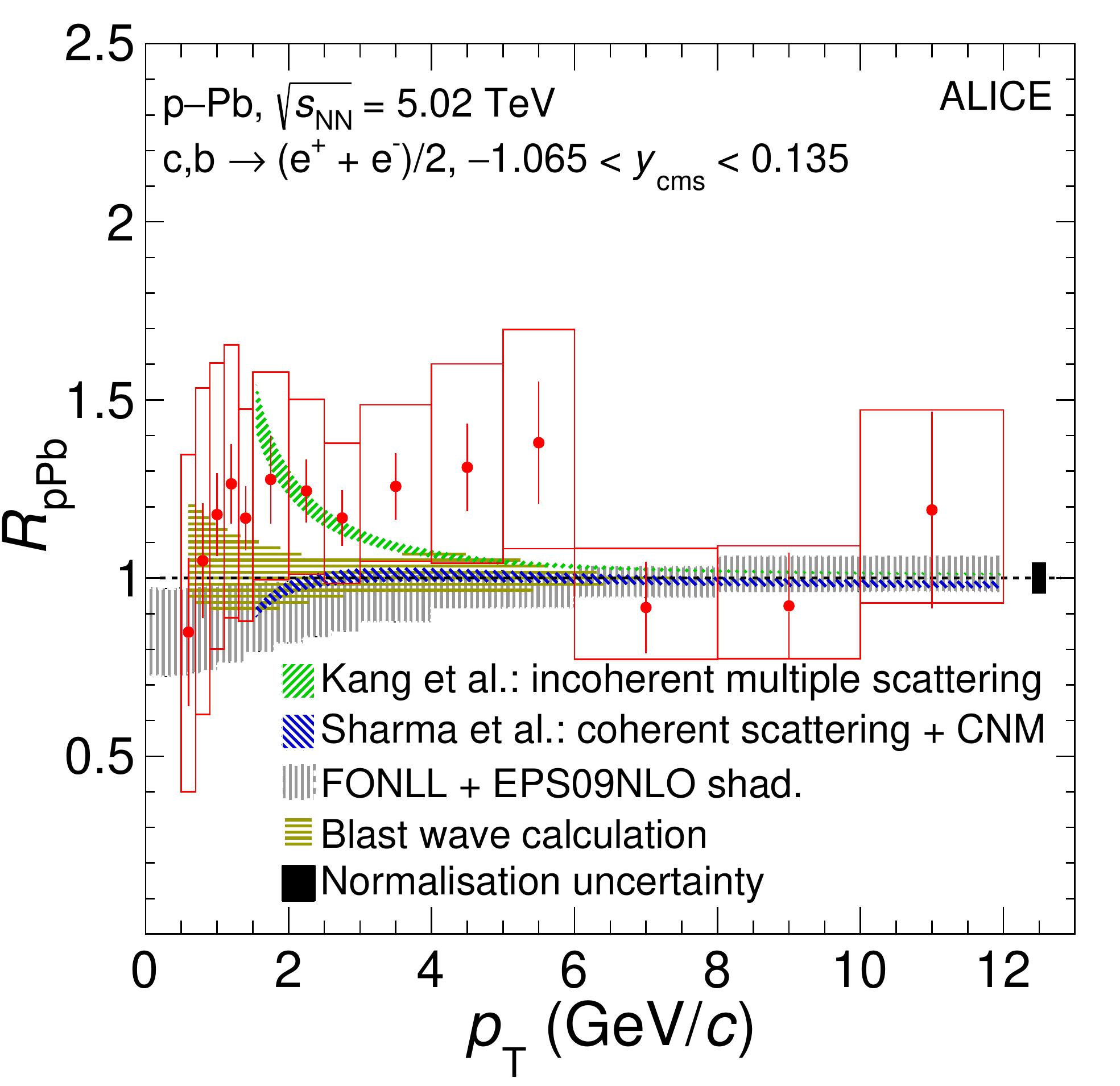}
  \caption{ Nuclear modification factor \RpPb of electrons from heavy-flavour hadron decays as a function of transverse momentum for minimum-bias \pPb collisions at \mbox{\sqpA}, compared with theoretical models~\cite{Cacciari:1998it,Eskola:2009uj,Sharma:2009hn,Kang:2014hha,Sickles:2013yna}, as described in the text. The vertical bars represent the statistical uncertainties, and the boxes indicate the systematic uncertainties. The systematic uncertainty from the  normalisation, common to all points, is shown as a filled box at high \pt.  
  }
\label{fig::RpPb}
 \end{figure}

The nuclear modification factor \RpPb of electrons from heavy-flavour hadron decays as a function of transverse momentum is shown in Fig.~\ref{fig::RpPb}. The statistical and systematic uncertainties of the spectra in \pPb and pp were propagated as independent uncertainties. The normalisation uncertainties of the pp reference and the \pPb spectrum were added in quadrature and are shown as a filled box at high transverse momentum in Fig.~\ref{fig::RpPb}. 

The \RpPb is consistent with unity within uncertainties over the whole \pt range of the measurement. The production of electrons from heavy-flavour hadron decays is thus consistent with binary collision scaling of the reference spectrum for pp collisions at the same centre-of-mass energy. The suppression of the yield of heavy-flavour production in Pb--Pb collisions at high-\pt is therefore a final state effect induced by the produced hot medium.

Given the large systematic uncertainties, our measurement is also compatible with an enhancement in the transverse momentum interval \ptrange{1}{6} as seen at mid-rapidity in d--Au collisions at $\sqrt{s_{{\rm NN}}} =$ 200~GeV~\cite{Adare:2012yxa}. Such an enhancement might be caused by radial flow as suggested by studies on the mean \pt as a function of the identified particle multiplicity~\cite{Abelev:2013haa}.

The data are described within the uncertainties by pQCD calculations including initial-state effects (FONLL~\cite{Cacciari:1998it} + \mbox{EPS09NLO~\cite{Eskola:2009uj}} nuclear shadowing parametrisation). The results suggest that initial-state effects are small at high transverse momentum in Pb--Pb collisions. Calculations by Sharma \textit{et al.} which include CNM energy loss, nuclear shadowing and coherent multiple scattering at the partonic level also describe the data~\cite{Sharma:2009hn}. Calculations based on incoherent multiple scatterings by Kang \textit{et al.} predict an enhancement at low \pt~\cite{Kang:2014hha}. The formation of a hydrodynamically expanding medium and consequently flow of charm and beauty quarks are expected to result in an enhancement in the nuclear modification factor \RpPb~\cite{Sickles:2013yna}. To quantify the possible effect on \RpPb, a blast wave calculation with parameters extracted from fits to the \pt spectra of light-flavour hadrons~\cite{Abelev:2013haa} measured in \pPb collisions was employed. The model calculation agrees with the data. However, the present uncertainties of the measurement do not allow us to discriminate among the aforementioned theoretical approaches.

\section{Summary and conclusions}\label{Sum}

The \pt-differential invariant cross section for electrons from heavy-flavour hadron decays in minimum-bias \pPb collisions at \mbox{\sqpA} was measured in the rapidity range $-1.065 < y_{\rm cms} < 0.135$ and the transverse momentum interval \ptrange{0.5}{12} using the combination of three electron identification methods. 
The application of the invariant mass technique to subtract electrons not originating from open heavy-flavour hadron decays largely reduced the systematic uncertainties with respect to the cocktail subtraction method, in particular at low transverse momentum. The pp reference for the nuclear modification factor \RpPb was obtained by interpolating the measured \pt-differential cross sections of electrons from heavy-flavour hadron decays at \s = 2.76~TeV and \s = 7~TeV. 
The \RpPb is consistent with unity within uncertainties of about 25\%, which become larger for \pt below 1~\GeVc. The presented calculations describe the data within uncertainties. The results suggest that heavy-flavour production in minimum-bias \pPb collisions scales with the number of binary collisions, although within uncertainties the data are also consistent with an enhancement above this scaling. The consistency with unity of the \RpPb at high \pt indicates that the suppression of heavy-flavour production in Pb--Pb collisions is of different origin than cold nuclear matter effects.

%%%%% acknowledgements
\newenvironment{acknowledgement}{\relax}{\relax}
\begin{acknowledgement}
\section*{Acknowledgements}
% $Id: acknowledgements.tex 2267 2015-09-07 08:16:22Z loizides $
% Version: Jan 2015

The ALICE Collaboration would like to thank all its engineers and technicians for their invaluable contributions to the construction of the experiment and the CERN accelerator teams for the outstanding performance of the LHC complex.
The ALICE Collaboration gratefully acknowledges the resources and support provided by all Grid centres and the Worldwide LHC Computing Grid (WLCG) Collaboration.
The ALICE Collaboration acknowledges the following funding agencies for their support in building and
running the ALICE detector:
State Committee of Science,  World Federation of Scientists (WFS)
and Swiss Fonds Kidagan, Armenia;
Conselho Nacional de Desenvolvimento Cient\'{\i}fico e Tecnol\'{o}gico (CNPq), Financiadora de Estudos e Projetos (FINEP),
Funda\c{c}\~{a}o de Amparo \`{a} Pesquisa do Estado de S\~{a}o Paulo (FAPESP);
National Natural Science Foundation of China (NSFC), the Chinese Ministry of Education (CMOE)
and the Ministry of Science and Technology of China (MSTC);
Ministry of Education and Youth of the Czech Republic;
Danish Natural Science Research Council, the Carlsberg Foundation and the Danish National Research Foundation;
The European Research Council under the European Community's Seventh Framework Programme;
Helsinki Institute of Physics and the Academy of Finland;
French CNRS-IN2P3, the `Region Pays de Loire', `Region Alsace', `Region Auvergne' and CEA, France;
German Bundesministerium fur Bildung, Wissenschaft, Forschung und Technologie (BMBF) and the Helmholtz Association;
General Secretariat for Research and Technology, Ministry of Development, Greece;
Hungarian Orszagos Tudomanyos Kutatasi Alappgrammok (OTKA) and National Office for Research and Technology (NKTH);
Department of Atomic Energy and Department of Science and Technology of the Government of India;
Istituto Nazionale di Fisica Nucleare (INFN) and Centro Fermi -
Museo Storico della Fisica e Centro Studi e Ricerche ``Enrico Fermi'', Italy;
MEXT Grant-in-Aid for Specially Promoted Research, Ja\-pan;
Joint Institute for Nuclear Research, Dubna;
National Research Foundation of Korea (NRF);
Consejo Nacional de Cienca y Tecnologia (CONACYT), Direccion General de Asuntos del Personal Academico(DGAPA), M\'{e}xico, Amerique Latine Formation academique - European Commission~(ALFA-EC) and the EPLANET Program~(European Particle Physics Latin American Network);
Stichting voor Fundamenteel Onderzoek der Materie (FOM) and the Nederlandse Organisatie voor Wetenschappelijk Onderzoek (NWO), Netherlands;
Research Council of Norway (NFR);
National Science Centre, Poland;
Ministry of National Education/Institute for Atomic Physics and National Council of Scientific Research in Higher Education~(CNCSI-UEFISCDI), Romania;
Ministry of Education and Science of Russian Federation, Russian
Academy of Sciences, Russian Federal Agency of Atomic Energy,
Russian Federal Agency for Science and Innovations and The Russian
Foundation for Basic Research;
Ministry of Education of Slovakia;
Department of Science and Technology, South Africa;
Centro de Investigaciones Energeticas, Medioambientales y Tecnologicas (CIEMAT), E-Infrastructure shared between Europe and Latin America (EELA), Ministerio de Econom\'{i}a y Competitividad (MINECO) of Spain, Xunta de Galicia (Conseller\'{\i}a de Educaci\'{o}n),
Centro de Aplicaciones Tecnológicas y Desarrollo Nuclear (CEA\-DEN), Cubaenerg\'{\i}a, Cuba, and IAEA (International Atomic Energy Agency);
Swedish Research Council (VR) and Knut $\&$ Alice Wallenberg
Foundation (KAW);
Ukraine Ministry of Education and Science;
United Kingdom Science and Technology Facilities Council (STFC);
The United States Department of Energy, the United States National
Science Foundation, the State of Texas, and the State of Ohio;
Ministry of Science, Education and Sports of Croatia and  Unity through Knowledge Fund, Croatia;
Council of Scientific and Industrial Research (CSIR), New Delhi, India;
Pontificia Universidad Cat\'{o}lica del Per\'{u}.
    %%%%%%% done by webmaster team
\end{acknowledgement}

%%%%%%%% Bibliography (In case of using bibtex generate the bbl requested by arXiv)
%\clearpage
\bibliographystyle{utphys}
\bibliography{hfe_pPb_mb.bib}

%%%%%%%%% appendix with author list
\newpage
\appendix
\section{The ALICE Collaboration}
\label{app:collab}

% Collaboration: CERN-LHC-ALICE
% Generation Date is 2015/Sep/16

% How to use:
%%%%%%%%% appendix with author list
%\appendix
%\section{The ALICE Collaboration}
%\label{app:collab}
%\input{authors-list.tex}  %%%%%%% get the latest version before submitting

\begingroup
\small
\begin{flushleft}
J.~Adam\Irefn{org40}\And
D.~Adamov\'{a}\Irefn{org83}\And
M.M.~Aggarwal\Irefn{org87}\And
G.~Aglieri Rinella\Irefn{org36}\And
M.~Agnello\Irefn{org110}\And
N.~Agrawal\Irefn{org48}\And
Z.~Ahammed\Irefn{org132}\And
S.U.~Ahn\Irefn{org68}\And
S.~Aiola\Irefn{org136}\And
A.~Akindinov\Irefn{org58}
% ...
\renewcommand\labelenumi{\textsuperscript{\theenumi}~}

\section*{Affiliation notes}
\renewcommand\theenumi{\roman{enumi}}
\begin{Authlist}
\item \Adef{0}Deceased
\item \Adef{idp1745584}{Also at: Georgia State University, Atlanta, Georgia, United States}
\item \Adef{idp3781840}{Also at: M.V. Lomonosov Moscow State University, D.V. Skobeltsyn Institute of Nuclear, Physics, Moscow, Russia}
\end{Authlist}

\section*{Collaboration Institutes}
\renewcommand\theenumi{\arabic{enumi}~}
\begin{Authlist}

\item \Idef{org1}A.I. Alikhanyan National Science Laboratory (Yerevan Physics Institute) Foundation, Yerevan, Armenia
\item \Idef{org2}Benem\'{e}rita Universidad Aut\'{o}noma de Puebla, Puebla, Mexico
\item \Idef{org3}Bogolyubov Institute for Theoretical Physics, Kiev, Ukraine
\item \Idef{org4}Bose Institute, Department of Physics and Centre for Astroparticle Physics and Space Science (CAPSS), Kolkata, India
\item \Idef{org5}Budker Institute for Nuclear Physics, Novosibirsk, Russia
\item \Idef{org6}California Polytechnic State University, San Luis Obispo, California, United States
\item \Idef{org7}Central China Normal University, Wuhan, China
\item \Idef{org8}Centre de Calcul de l'IN2P3, Villeurbanne, France
\item \Idef{org9}Centro de Aplicaciones Tecnol\'{o}gicas y Desarrollo Nuclear (CEADEN), Havana, Cuba
\item \Idef{org10}Centro de Investigaciones Energ\'{e}ticas Medioambientales y Tecnol\'{o}gicas (CIEMAT), Madrid, Spain
\item \Idef{org11}Centro de Investigaci\'{o}n y de Estudios Avanzados (CINVESTAV), Mexico City and M\'{e}rida, Mexico
\item \Idef{org12}Centro Fermi - Museo Storico della Fisica e Centro Studi e Ricerche ``Enrico Fermi'', Rome, Italy
\item \Idef{org13}Chicago State University, Chicago, Illinois, USA
\item \Idef{org14}China Institute of Atomic Energy, Beijing, China
\item \Idef{org15}Commissariat \`{a} l'Energie Atomique, IRFU, Saclay, France
\item \Idef{org16}COMSATS Institute of Information Technology (CIIT), Islamabad, Pakistan
\item \Idef{org17}Departamento de F\'{\i}sica de Part\'{\i}culas and IGFAE, Universidad de Santiago de Compostela, Santiago de Compostela, Spain
\item \Idef{org18}Department of Physics and Technology, University of Bergen, Bergen, Norway
\item \Idef{org19}Department of Physics, Aligarh Muslim University, Aligarh, India
\item \Idef{org20}Department of Physics, Ohio State University, Columbus, Ohio, United States
\item \Idef{org21}Department of Physics, Sejong University, Seoul, South Korea
\item \Idef{org22}Department of Physics, University of Oslo, Oslo, Norway
\item \Idef{org23}Dipartimento di Elettrotecnica ed Elettronica del Politecnico, Bari, Italy
\item \Idef{org24}Dipartimento di Fisica dell'Universit\`{a} 'La Sapienza' and Sezione INFN Rome, Italy
\item \Idef{org25}Dipartimento di Fisica dell'Universit\`{a} and Sezione INFN, Cagliari, Italy
\item \Idef{org26}Dipartimento di Fisica dell'Universit\`{a} and Sezione INFN, Trieste, Italy
\item \Idef{org27}Dipartimento di Fisica dell'Universit\`{a} and Sezione INFN, Turin, Italy
\item \Idef{org28}Dipartimento di Fisica e Astronomia dell'Universit\`{a} and Sezione INFN, Bologna, Italy
\item \Idef{org29}Dipartimento di Fisica e Astronomia dell'Universit\`{a} and Sezione INFN, Catania, Italy
\item \Idef{org30}Dipartimento di Fisica e Astronomia dell'Universit\`{a} and Sezione INFN, Padova, Italy
\item \Idef{org31}Dipartimento di Fisica `E.R.~Caianiello' dell'Universit\`{a} and Gruppo Collegato INFN, Salerno, Italy
\item \Idef{org32}Dipartimento di Scienze e Innovazione Tecnologica dell'Universit\`{a} del  Piemonte Orientale and Gruppo Collegato INFN, Alessandria, Italy
\item \Idef{org33}Dipartimento Interateneo di Fisica `M.~Merlin' and Sezione INFN, Bari, Italy
\item \Idef{org34}Division of Experimental High Energy Physics, University of Lund, Lund, Sweden
\item \Idef{org35}Eberhard Karls Universit\"{a}t T\"{u}bingen, T\"{u}bingen, Germany
\item \Idef{org36}European Organization for Nuclear Research (CERN), Geneva, Switzerland
\item \Idef{org37}Excellence Cluster Universe, Technische Universit\"{a}t M\"{u}nchen, Munich, Germany
\item \Idef{org38}Faculty of Engineering, Bergen University College, Bergen, Norway
\item \Idef{org39}Faculty of Mathematics, Physics and Informatics, Comenius University, Bratislava, Slovakia
\item \Idef{org40}Faculty of Nuclear Sciences and Physical Engineering, Czech Technical University in Prague, Prague, Czech Republic
\item \Idef{org41}Faculty of Science, P.J.~\v{S}af\'{a}rik University, Ko\v{s}ice, Slovakia
\item \Idef{org42}Faculty of Technology, Buskerud and Vestfold University College, Vestfold, Norway
\item \Idef{org43}Frankfurt Institute for Advanced Studies, Johann Wolfgang Goethe-Universit\"{a}t Frankfurt, Frankfurt, Germany
\item \Idef{org44}Gangneung-Wonju National University, Gangneung, South Korea
\item \Idef{org45}Gauhati University, Department of Physics, Guwahati, India
\item \Idef{org46}Helsinki Institute of Physics (HIP), Helsinki, Finland
\item \Idef{org47}Hiroshima University, Hiroshima, Japan
\item \Idef{org48}Indian Institute of Technology Bombay (IIT), Mumbai, India
\item \Idef{org49}Indian Institute of Technology Indore, Indore (IITI), India
\item \Idef{org50}Inha University, Incheon, South Korea
\item \Idef{org51}Institut de Physique Nucl\'eaire d'Orsay (IPNO), Universit\'e Paris-Sud, CNRS-IN2P3, Orsay, France
\item \Idef{org52}Institut f\"{u}r Informatik, Johann Wolfgang Goethe-Universit\"{a}t Frankfurt, Frankfurt, Germany
\item \Idef{org53}Institut f\"{u}r Kernphysik, Johann Wolfgang Goethe-Universit\"{a}t Frankfurt, Frankfurt, Germany
\item \Idef{org54}Institut f\"{u}r Kernphysik, Westf\"{a}lische Wilhelms-Universit\"{a}t M\"{u}nster, M\"{u}nster, Germany
\item \Idef{org55}Institut Pluridisciplinaire Hubert Curien (IPHC), Universit\'{e} de Strasbourg, CNRS-IN2P3, Strasbourg, France
\item \Idef{org56}Institute for Nuclear Research, Academy of Sciences, Moscow, Russia
\item \Idef{org57}Institute for Subatomic Physics of Utrecht University, Utrecht, Netherlands
\item \Idef{org58}Institute for Theoretical and Experimental Physics, Moscow, Russia
\item \Idef{org59}Institute of Experimental Physics, Slovak Academy of Sciences, Ko\v{s}ice, Slovakia
\item \Idef{org60}Institute of Physics, Academy of Sciences of the Czech Republic, Prague, Czech Republic
\item \Idef{org61}Institute of Physics, Bhubaneswar, India
\item \Idef{org62}Institute of Space Science (ISS), Bucharest, Romania
\item \Idef{org63}Instituto de Ciencias Nucleares, Universidad Nacional Aut\'{o}noma de M\'{e}xico, Mexico City, Mexico
\item \Idef{org64}Instituto de F\'{\i}sica, Universidad Nacional Aut\'{o}noma de M\'{e}xico, Mexico City, Mexico
\item \Idef{org65}iThemba LABS, National Research Foundation, Somerset West, South Africa
\item \Idef{org66}Joint Institute for Nuclear Research (JINR), Dubna, Russia
\item \Idef{org67}Konkuk University, Seoul, South Korea
\item \Idef{org68}Korea Institute of Science and Technology Information, Daejeon, South Korea
\item \Idef{org69}KTO Karatay University, Konya, Turkey
\item \Idef{org70}Laboratoire de Physique Corpusculaire (LPC), Clermont Universit\'{e}, Universit\'{e} Blaise Pascal, CNRS--IN2P3, Clermont-Ferrand, France
\item \Idef{org71}Laboratoire de Physique Subatomique et de Cosmologie, Universit\'{e} Grenoble-Alpes, CNRS-IN2P3, Grenoble, France
\item \Idef{org72}Laboratori Nazionali di Frascati, INFN, Frascati, Italy
\item \Idef{org73}Laboratori Nazionali di Legnaro, INFN, Legnaro, Italy
\item \Idef{org74}Lawrence Berkeley National Laboratory, Berkeley, California, United States
\item \Idef{org75}Moscow Engineering Physics Institute, Moscow, Russia
\item \Idef{org76}Nagasaki Institute of Applied Science, Nagasaki, Japan
\item \Idef{org77}National Centre for Nuclear Studies, Warsaw, Poland
\item \Idef{org78}National Institute for Physics and Nuclear Engineering, Bucharest, Romania
\item \Idef{org79}National Institute of Science Education and Research, Bhubaneswar, India
\item \Idef{org80}Niels Bohr Institute, University of Copenhagen, Copenhagen, Denmark
\item \Idef{org81}Nikhef, Nationaal instituut voor subatomaire fysica, Amsterdam, Netherlands
\item \Idef{org82}Nuclear Physics Group, STFC Daresbury Laboratory, Daresbury, United Kingdom
\item \Idef{org83}Nuclear Physics Institute, Academy of Sciences of the Czech Republic, \v{R}e\v{z} u Prahy, Czech Republic
\item \Idef{org84}Oak Ridge National Laboratory, Oak Ridge, Tennessee, United States
\item \Idef{org85}Petersburg Nuclear Physics Institute, Gatchina, Russia
\item \Idef{org86}Physics Department, Creighton University, Omaha, Nebraska, United States
\item \Idef{org87}Physics Department, Panjab University, Chandigarh, India
\item \Idef{org88}Physics Department, University of Athens, Athens, Greece
\item \Idef{org89}Physics Department, University of Cape Town, Cape Town, South Africa
\item \Idef{org90}Physics Department, University of Jammu, Jammu, India
\item \Idef{org91}Physics Department, University of Rajasthan, Jaipur, India
\item \Idef{org92}Physik Department, Technische Universit\"{a}t M\"{u}nchen, Munich, Germany
\item \Idef{org93}Physikalisches Institut, Ruprecht-Karls-Universit\"{a}t Heidelberg, Heidelberg, Germany
\item \Idef{org94}Purdue University, West Lafayette, Indiana, United States
\item \Idef{org95}Pusan National University, Pusan, South Korea
\item \Idef{org96}Research Division and ExtreMe Matter Institute EMMI, GSI Helmholtzzentrum f\"ur Schwerionenforschung, Darmstadt, Germany
\item \Idef{org97}Rudjer Bo\v{s}kovi\'{c} Institute, Zagreb, Croatia
\item \Idef{org98}Russian Federal Nuclear Center (VNIIEF), Sarov, Russia
\item \Idef{org99}Russian Research Centre Kurchatov Institute, Moscow, Russia
\item \Idef{org100}Saha Institute of Nuclear Physics, Kolkata, India
\item \Idef{org101}School of Physics and Astronomy, University of Birmingham, Birmingham, United Kingdom
\item \Idef{org102}Secci\'{o}n F\'{\i}sica, Departamento de Ciencias, Pontificia Universidad Cat\'{o}lica del Per\'{u}, Lima, Peru
\item \Idef{org103}Sezione INFN, Bari, Italy
\item \Idef{org104}Sezione INFN, Bologna, Italy
\item \Idef{org105}Sezione INFN, Cagliari, Italy
\item \Idef{org106}Sezione INFN, Catania, Italy
\item \Idef{org107}Sezione INFN, Padova, Italy
\item \Idef{org108}Sezione INFN, Rome, Italy
\item \Idef{org109}Sezione INFN, Trieste, Italy
\item \Idef{org110}Sezione INFN, Turin, Italy
\item \Idef{org111}SSC IHEP of NRC Kurchatov institute, Protvino, Russia
\item \Idef{org112}Stefan Meyer Institut f\"{u}r Subatomare Physik (SMI), Vienna, Austria
\item \Idef{org113}SUBATECH, Ecole des Mines de Nantes, Universit\'{e} de Nantes, CNRS-IN2P3, Nantes, France
\item \Idef{org114}Suranaree University of Technology, Nakhon Ratchasima, Thailand
\item \Idef{org115}Technical University of Ko\v{s}ice, Ko\v{s}ice, Slovakia
\item \Idef{org116}Technical University of Split FESB, Split, Croatia
\item \Idef{org117}The Henryk Niewodniczanski Institute of Nuclear Physics, Polish Academy of Sciences, Cracow, Poland
\item \Idef{org118}The University of Texas at Austin, Physics Department, Austin, Texas, USA
\item \Idef{org119}Universidad Aut\'{o}noma de Sinaloa, Culiac\'{a}n, Mexico
\item \Idef{org120}Universidade de S\~{a}o Paulo (USP), S\~{a}o Paulo, Brazil
\item \Idef{org121}Universidade Estadual de Campinas (UNICAMP), Campinas, Brazil
\item \Idef{org122}University of Houston, Houston, Texas, United States
\item \Idef{org123}University of Jyv\"{a}skyl\"{a}, Jyv\"{a}skyl\"{a}, Finland
\item \Idef{org124}University of Liverpool, Liverpool, United Kingdom
\item \Idef{org125}University of Tennessee, Knoxville, Tennessee, United States
\item \Idef{org126}University of the Witwatersrand, Johannesburg, South Africa
\item \Idef{org127}University of Tokyo, Tokyo, Japan
\item \Idef{org128}University of Tsukuba, Tsukuba, Japan
\item \Idef{org129}University of Zagreb, Zagreb, Croatia
\item \Idef{org130}Universit\'{e} de Lyon, Universit\'{e} Lyon 1, CNRS/IN2P3, IPN-Lyon, Villeurbanne, France
\item \Idef{org131}V.~Fock Institute for Physics, St. Petersburg State University, St. Petersburg, Russia
\item \Idef{org132}Variable Energy Cyclotron Centre, Kolkata, India
\item \Idef{org133}Warsaw University of Technology, Warsaw, Poland
\item \Idef{org134}Wayne State University, Detroit, Michigan, United States
\item \Idef{org135}Wigner Research Centre for Physics, Hungarian Academy of Sciences, Budapest, Hungary
\item \Idef{org136}Yale University, New Haven, Connecticut, United States
\item \Idef{org137}Yonsei University, Seoul, South Korea
\item \Idef{org138}Zentrum f\"{u}r Technologietransfer und Telekommunikation (ZTT), Fachhochschule Worms, Worms, Germany
\end{Authlist}
\endgroup

  %%%%%%% done by webmaster team
\end{document}